\documentclass[a4paper]{article}

\usepackage[english]{babel}
\usepackage[utf8x]{inputenc}
\usepackage{csquotes}
\usepackage{soul}
\usepackage[T1]{fontenc}
\usepackage{enumitem}
\usepackage{subcaption}
\usepackage{parskip}
\usepackage{eufrak}
\usepackage{braket}
\usepackage[a4paper,top=3cm,bottom=2cm,left=3cm,right=3cm,marginparwidth=1.75cm]{geometry}
\usepackage[affil-it]{authblk}
\usepackage{amsmath}
\usepackage{amsthm}
\usepackage{amssymb}
\usepackage{amsfonts}
\usepackage{mathtools}
\usepackage{graphicx}
\usepackage[table,xcdraw]{xcolor}
\usepackage[colorinlistoftodos]{todonotes}
\usepackage[colorlinks=true, allcolors=blue]{hyperref}
\title{Heisenberg's Equality of Inequivalents Problem}
\author{Armin Nikkhah Shirazi \thanks{armin@umich.edu}}
\affil{\large University of Michigan, Ann Arbor}
\date{}
\begin{document}
\maketitle

\section{Introduction}

In the academic year 1954-1955, Werner Heisenberg delivered the Griffin lectures at the University of St. Andrews entitled \emph{Physics and Philosophy: The Revolution in Modern Science}. He subsequently published the lectures in a book of the same title (Heisenberg, 1958). In it, he expressed a distinction, inspired by the philosophy of Aristotle, which is encapsulated by the following passage\footnote{Heisenberg was not the first to make this proposal. For example, in 1940 the German Philosopher and diplomat Kurt Riezler expressed (in the voice of Aristotle) a similar view  in his lecture series and subsequent book \emph{Physics and Reality} (Riezler, 1940).
However, for Riezler, the  Aristotelian dichotomy between potentiality and actuality was just one of a set of dichotomies, some of which, such as that between mind and matter, body and soul, and so on went beyond the domain of physics. Heisenberg, on the other hand, stayed much more within the confines of physics, and for that reason I will continue to refer to the distinction as Heisenberg's.}: 
\begin{quotation}
\noindent In the experiments about atomic events we have to do with things and facts, with phenomena that are just as real as any phenomena in daily life. But the atoms or the elementary particles themselves are not as real; they form a world of potentialities or possibilities rather than of things or facts.(p. 160)
\end{quotation}
Heisenberg's distinction, naturally interpreted in ontological terms\footnote{It should be noted that nearly $25$ years prior to publication of this text, Heisenberg's view of the dualism seemed to be much more epistemic and similar to Bohr's:
\begin{quotation}
    ``An analysis of that situation we owe in the first place to Bohr, and this can only be indicated here. It appears that a peculiar schism in our investigations of atomic processes is inevitable. On the one hand the experimental questions which we ask of nature are always formulated with the help of the plain concepts of classical physics and more especially using the concepts of time and space.[...] On the other hand, the mathematical expressions suitable for the representation of experimental reality are wave functions in multi-dimensional configuration space which allow of no easily comprehensible interpretation. Out of this schism there arises the necessity to draw a clear dividing line in the description of atomic processes, between the measured apparatus of the observer which is described by classical concepts, and the object under observation, whose behavior is represented by a wave function.'' (Heisenberg, 1954)
    \end{quotation}
     (This text is part of a speech originally delivered at the first general session on the occasion of the General Meeting of the ``Gesellschaft Deutscher Naturforscher and Aerzte,''  Hanover, on 17th September, 1934)} , seems like a rather attractive option to gain intuition about quantum mechanics, and his view has over the years been echoed by others (Fock, 1957; Shimony, 1990; Kastner, Kauffmann and Epperson, 2017). However, it turns out that trying to implement the distinction in a serious manner into the mathematical formalism of quantum mechanics results in a conceptual problem that renders the formalism incoherent.\\
In this paper I will show how this problem, which I call the \emph{equality of inequivalents problem}, arises and then argue that this problem should not be interpreted as a reason for discarding Heisenberg's distinction, but rather as a reason to consider enriching the formalism of quantum mechanics so that it can accommodate the distinction. My argument will be by analogy: I identify a similar problem in axiomatic probability and then propose a resolution by means of an enriched axiomatization.\\

\section{``Potentialities or Possibilities'' vs. ``Things or Facts''}
If we wish to implement Heisenberg's distinction into the mathematical formalism of quantum mechanics, we should first be clear what the difference is between a ``potentiality or possibility'' and a ``thing or fact''. While most of us have good intuitions which allow us to easily distinguish them from each other, that is not enough. We should be able to ground the difference in a clearly articulated characteristic common to members of each family that sets them apart from members of the other.\\
Whatever other properties they may have, it seems in accord with our intuitions that incompatible ``potentialities or possibilities'' can coexist, while incompatible ``things or facts'' cannot. It may well be that Heisenberg had precisely this distinction in mind, given that the central feature of quantum mechanics, namely quantum superposition, can be fairly described as the coexistence of incompatible physical states.\\
We will, therefore, take this to be the key criterion to  distinguish between the two families. Now, this distinction is sufficiently strong that it eliminates any overlap between them: the possibility of incompatibilities immediately eliminates a candidate for a fact from consideration as such, while the impossibility of incompatibilities \emph{apart from individual circumstances} locks a candidate for a fact in as such. The qualifier ``apart from individual circumstances'' is necessary in order to account for situations in which individual circumstances have been contrived so that only a single possibility is available. For example, if a die is rigged so that it is certain that only one outcome is possible \emph{if it is thrown}, then the unavailability of other possibilities still does not render this certainty the same as the situation in which the die has actually been thrown, the outcome of which we would consider a fact.  The possibility appears only ``fact-like'' here because the individual circumstances have been contrived to make it so. In general, the outcome of a throw of a die is open to more than a single possibility, all of which are incompatible, and it is this more general perspective I have in mind when distinguishing the two families with respect to whether they accommodate the coexistence of incompatibilities. I believe this comports with our intuitions, according to which the certainty of an outcome not yet obtained does not make it the same as an outcome that has been obtained. \\
Beside the absence of an overlap between the two families, we need consider one other feature: Every physical system that we take to be in the world can be described in terms of either ``potentialities or possibilities'' or ``things or facts''. Of course, we can always imagine physical systems which are impossible, as is often done in the science fiction and fantasy literature, but we would not take such systems to ``be'' in the world, except under some worldviews which confer being to things regardless of whether they can be found in reality or not. Setting such worldviews aside, it seems safe to assume that the two families `cover' the set of all descriptions of physical states that can be found. \\
If both of these properties hold, namely, that ``potentialities or possibilities'' and ``things or facts'' are mutually exclusive, and that they cover the entire set of descriptions of physical states, then that implies that they \emph{partition} that set. This, in turn, implies that these two families are really \emph{equivalence classes}.  This will be a key feature which causes a problem.

\section{The Equality of Inequivalents Problem}
Consider the quantum superposition equation:
    \begin{equation}
       \bra \psi = \sum\limits_{k} c_k \bra {\psi_k}
    \end{equation}
   In a strictly mathematical sense, it says that a quantum state, regarded as an element of Hilbert space, can be expressed in terms of a combination or superposition of other elements under a change of basis (i.e. a change of a set of linearly independent vectors which span the space).\\
   In a minimalist physical sense, it seems to say that any given quantum state is in the strictest sense equivalent to a combination of other physical states if one changes the property of the state one wishes to ``measure''. The difficulty of reconciling this minimalist physical interpretation with our immediate experience of the world can be regarded as one of the reasons for the proliferation of interpretations of quantum mechanics.\\
   If we now apply Heisenberg's distinction to this equation, then it permits at least two distinct interpretations, depending on the circumstances the equation is meant to model. One interpretation is unproblematic and the other is problematic. \begin{itemize}
        \item Interpretation $1$:  The right side represents a combination of coexisting incompatible states, so by the criterion discussed in the previous section, it must represent a combination of ``potentialities or possibilities''. The left side represents an unmeasured quantum state. Unmeasured quantum states are precisely what Heisenberg referred to as ``potentialities or possibilities'', so this interpretation is unproblematic, as both sides of the equality belong to the same equivalence class. 
        \item Interpretation $2$: As before, the right side must be considered a combination of ``potentialities or possibilities'' because it expresses coexisting incompatible states, but the left side represents a quantum state that is under measurement. We have to consider this to represent a ``thing or fact'' since, when Heisenberg said  ``In experiments about atomic events, we have to do with things and facts...'', he was assigning results of experiments with and measurements on quantum systems to this family. This is problematic because the two sides belong to different equivalence classes, yet there is an equality between them\footnote{A third interpretation, that the left side represents a ``potentiality or possibility'' while the right side represents a combination of ``things or facts'' is harder to make sense of, but at any rate suffers the same problem as the second interpretation. A fourth interpretation, that both sides represent ``things or facts'', does not suffer this problem and is in fact licensed by Koopman-von Neumann theory, a Hilbert space-based formulation of classical mechanics(Koopman, 1931; von Neumann, 1932). However, because classical Hilbert spaces lack the essential features which give rise to the problems Heisenberg sought to address with his distinction, such as non-commuting operators, the fourth interpretation is not relevant to the present discussion.}.
    \end{itemize}
    The fact that the formalism of quantum mechanics permits the second interpretation means that if one attempts to implement Heisenberg's distinction into it, then this renders the entire formalism incoherent. This is in a similar sense as, say, number theory would become incoherent if we attempted to set a member of the equivalence class of even numbers equal to a member of the equivalence class of odd numbers. I call this the \emph{equality of inequivalents problem}.

\section{Three Possible Reactions}
There are three possible ways to react to the equality of inequivalents problem: 
\begin{enumerate}
    \item \textbf{Deny the problem by challenging some of the assumptions that underlie its formulation.} It is essential to the problem that ``potentialities and possibilities'' and ``things or facts'' be regarded as ontological equivalence classes. Is there no way it could be otherwise? One strategy would be to take the distinction to be purely epistemological, which would rob it of its power as a claim about a distinction ``in the real world out there''. Niels Bohr, who made a superficially similar classical-quantum distinction, seems to have regarded it this way (Bohr, 1928). In Bohr's view, the distinction was not the result of any division in reality, but rather a manifestation of our inability to grasp the quantum world directly: the only concepts at our disposal, which are classical, are inadequate, but we can establish an epistemological correspondence between our classical concepts and the quantum world through such a distinction.  As a result, there is no guarantee that Bohr's distinction is related to an ontological equivalence class structure and it is actually far more plausible that it is not.\\
    A second strategy would be to deny the key criterion for distinguishing between the two families. But in order for this approach to succeed, it must affirmatively propose a suitable alternative criterion which $(a)$ avoids the ontological equivalence class structure and $(b)$ reflects our intuitions about the distinctions more closely. As it seems quite in agreement with our intuitions that Heisenberg's distinction is really one between two ontological equivalence relations, one may be skeptical whether such a candidate criterion exists. \\
    A third strategy would be to take on a general worldview in which the equivalence class structure becomes untenable. For instance, a worldview in which each possibility is as real as a fact, such as David Lewis' modal realism (Lewis, 1973), would altogether annihilate the equivalence class distinction underlying the argument.\\
    These strategies seem to either change Heisenberg's distinction into something else, or otherwise require us to adopt worldviews which are very different than what most might agree on. This renders them, at least for the purposes of this paper, less interesting and so we will move on to considering other possible reactions. 
    \item \textbf{Accept the problem and take it to be grounds for dismissing Heisenberg's or similar distinctions altogether.} Even apart from a discussion of Heisenberg's distinction, it seems to be the view of many physicists that the quantum-classical distinction is meaningless. The advantage of a reaction along these lines is that it finds solid grounding in the mathematical formalism of quantum mechanics, as indeed the equality of inequivalents problem shows. However, that very advantage also presents a risk. As Korzybski said, ``the map is not the territory'', by which he meant that we should take care to never conflate our models of reality with reality itself (Korzybski, 1994). So even if our model of reality suggests that a certain distinction is meaningless, there is still at least a possibility that this is not because the distinction actually is so, but because the model inadequately represents reality with respect to that distinction. If this is indeed the case, then our dismissal of the distinction would mean that we have conflated the map with the territory. The last reaction is based on taking this possibility seriously.
    \item \textbf{Accept the problem but take its source to be not Heisenberg's distinction but the quantum formalism.} The mathematical features which give rise to the problem are that $(a)$ the Hilbert space contains all physical states as elements, and $(b)$ the structure of this space allows any element to be expressed in terms of a quantum superposition. This makes it inevitable that even states under measurement are expressible in terms of a quantum superposition.   But is this combination of mathematical features absolutely unavoidable? Can we, for instance, not come up with a richer mathematical structure which respects the distinction, yet under their denial collapses to the standard formalism?  It may certainly seem presumptuous to want to modify a time-honored mathematical formalism, but it will seem less so if a similar problem can be identified elsewhere and resolved by just this strategy. 
\end{enumerate}

As I am partial to the third reaction, I will now attempt to demonstrate that a version of the equality of inequivalents problem can be found and resolved in axiomatic probability theory.

\section{Enriching Axiomatic Probability}
A striking feature of axiomatic probability (Kolmogorov, 1950) is that we cannot distinguish, \emph{within the mathematics}, between a probability and a non-probabilistic unit measure, such as a unit length. In other words, the formalism of probability does not permit us to distinguish between the concept of probability as a unit measure on a set of possibilities and the concept of a unit measure on things that are not possibilities. Here we should note that the \emph{concept} of probability is independent of its \emph{interpretation}: whether the probability is interpreted as ontic (e.g. frequentist or dispositionalist) or epistemic (e.g. Bayesian), in either case one can meaningfully distinguish between possibilities and things which are not possibilities, such as regions of a unit stick.\\
So, if we used words, as Heisenberg did, to attempt to distinguish between measures on possibilities and measures on things which are not possibilities \emph{within the mathematics} in a mathematical formalism that treats them exactly the same, then we run once again into an equality of inequivalents.\\
As a matter of practice, the problem is hardly noticed, for two reasons: first, whereas quantum mechanics is a theory in physics, probability is a theory in mathematics. Consequently,  questions about the ontological status of the objects in their domains are far less pressing for the latter. Second, our mathematical models have (so far!) never required us to use the \emph{concept} of probability when doing calculations. If needed, we could always from the context of our application-\emph{and as a strictly extramathematical matter}-conveniently distinguish between a probability and a non-probabilistic unit measure.\\
However, in the present discussion we consider it important that the mathematical formalism capture the concept, and I claim that it is not at all difficult to modify the formalism so that it can do so. I will prove my claim by construction. In the following, the standard axiomatization of probability is enriched by extra structure meant to help mathematically capture the concept of probability: 
\begin{quotation}
\noindent Let $\Omega=\bigcup_{i=1}^{N}E_i$ be a set where $N$ is either finite or countably infinite, $\mathcal{A} \subseteq \mathcal{P}(\Omega)$ a set of its mutually exclusive subsets $E_i$, and call the pair $(\Omega, \mathcal{A})$ a measurable space. Let $\Gamma=\{\gamma|\gamma=f(\omega)\}$ be a set where $f: \Omega \rightarrow \Gamma $ is a bijection and let $g:\Omega \rightarrow \Gamma$ be another map. A real-valued function $P:\mathcal{A} \rightarrow \mathbb{R}$ satisfying 
\begin{itemize}

\item{{Axiom $0$: $P(g^{-1}(\gamma))=P(\Omega)$}}
\color{black}
\item{Axiom $1$: $P(\Omega)=1$}
\item{Axiom $2$: $0 \leq P(E_i) \leq 1$ }

\item{Axiom $3$: $P\bigcup_{i=1}^{N}E_i=\sum_{i=1 }^{N}P(E_i)$ }
\end{itemize}
is called a \emph{probability}.
\end{quotation}
The additional structure here consists of:
\begin{enumerate}
    \item a set $\Gamma$, called the \emph{outcome set}, which represents the collection of ``things and facts'', to be distinguished from ``possibilities and potentialities'' which are represented by the sample space $\Omega$.
    \item a map $f$, called the \emph{correspondence function}, which brings each element of the sample space, into a correspondence with exactly one element of the outcome set.
    \item an axiom which says that probability is a measure over a fiber $g^{-1}$ on each element of the outcome set such that this fiber is just what we call the sample space.
\end{enumerate}

Calculationally, this axiomatization is equivalent to the standard axiomatization, but unlike the latter, it actually captures the concept of probability. To illustrate this, consider the example of a die roll: before rolling the die, the outcomes exist only as ``potentialities or possibilities''. Mathematically, the collection of these outcomes is modeled by a fiber on the outcome that will result from a throw. The probability is then defined as a measure on this fiber, but not on the outcome set. Hence, if, for example a die roll produces an outcome of $3$, say, then this outcome as an element of the fiber has a probability associated with it, but as an element of the outcome set it does not: it is simply a ``thing or fact''.\\
A formalization of quantum mechanics is in preparation which takes an approach analogous to the approach taken here to mathematically implement Heisenberg's distinction (Nikkhah Shirazi). 

\pagebreak
%\bibliography{Heisenberg}
%\bibliographystyle{plain}

\end{document}